\documentclass[preprint,12pt]{elsarticle}

\usepackage{amssymb}
\usepackage{array,booktabs,multirow,longtable,makecell}

\journal{Discover Applied Sciences}

\begin{document}

\begin{frontmatter}
\date{}


\title{Constrained Adversarial Learning for Automated Software Testing: a literature review}


\author[isep,gecad]{Jo{\~{a}}o Vitorino\corref{c1}}
\ead{jpmvo@isep.ipp.pt}

\author[isep,gecad]{Tiago Dias}
\ead{tiada@isep.ipp.pt}

\author[isep]{Tiago Fonseca}
\ead{calof@isep.ipp.pt}

\author[isep,gecad]{Eva Maia}
\ead{egm@isep.ipp.pt}

\author[isep,gecad]{Isabel Pra{\c{c}}a}
\ead{icp@isep.ipp.pt}

\cortext[c1]{Corresponding author}

% \address[isep]{{School of Engineering, Polytechnic of Porto (ISEP/IPP)}, {4249-015}, {Porto}, {Portugal}}

% \address[gecad]{{Research Group on Intelligent Engineering and Computing for Advanced Innovation and Development (GECAD)}, {4249-015}, {Porto}, {Portugal}}

\affiliation[isep]{organization={ISEP, Polytechnic of Porto},
            postcode={4249-015},
            city={Porto},
            country={Portugal}}

\affiliation[gecad]{organization={Research Group on Intelligent Engineering and Computing for Advanced Innovation and Development (GECAD)},
            postcode={4249-015},
            city={Porto},
            country={Portugal}}

\begin{abstract}
It is imperative to safeguard computer applications and information systems against the growing number of cyber-attacks. Automated software testing tools can be developed to quickly analyze many lines of code and detect vulnerabilities by generating function-specific testing data. This process draws similarities to the constrained adversarial examples generated by adversarial machine learning methods, so there could be significant benefits to the integration of these methods in testing tools to identify possible attack vectors. Therefore, this literature review is focused on the current state-of-the-art of constrained data generation approaches applied for adversarial learning and software testing, aiming to guide researchers and developers to enhance their software testing tools with adversarial testing methods and improve the resilience and robustness of their information systems. The found approaches were systematized, and the advantages and limitations of those specific for white-box, grey-box, and black-box testing were analyzed, identifying research gaps and opportunities to automate the testing tools with data generated by adversarial attacks.
\end{abstract}

%% \begin{graphicalabstract}
%% \includegraphics{grabs}
%% \end{graphicalabstract}

%% \begin{highlights}
%% \item Research highlight 1
%% \item Research highlight 2
%% \end{highlights}


\end{frontmatter}

%% This is file 'section-introduction.tex'
\section{Introduction}
\label{sec:intro}

From decision-making processes to electronic commerce and a variety of online services, the dependency of modern organizations on computer programs and information systems is not slowing down soon \cite{EuropeanCommission2021}. However, no computer code is flawless, so the integration of different technologies and software components can add hidden vulnerabilities ready to be exploited in novel zero-day cyber-attacks. Not even systems that rely on artificial intelligence are secure, due to the susceptibility of machine learning models to attacks with adversarial examples \cite{Vitorino2023}. As organizations rush the deployment of computer applications without thoroughly testing all software components, they are increasing the risk of an attacker with malicious intents disrupting their critical business processes and impacting their business continuity \cite{EuropeanUnionAgencyforCybersecurity2022}.

To safeguard organizations and their personnel from the damage caused by an attack, it is imperative to integrate security best practices in software development workflows and ensure that all computer applications are adequately tested before being deployed. Nonetheless, despite growing concerns for software testing, it is still mostly a manual process where developers design the tests that each distinct part of the code must be validated against. This process is time-consuming, expensive, and usually incomplete, because human personnel cannot formulate tests for the entire attack surface of an information system nor tackle all possible attack vectors \cite{EuropeanUnionAgencyforCybersecurity2022a}.

Automated software testing is a very promising research field because an automated and methodical approach can address the attack vectors that would otherwise be left unnoticed. By quickly analyzing thousands of lines of code, both time and monetary costs can be reduced. Nonetheless, to perform trustworthy tests and ensure that a program is functioning correctly, the generated testing data must address the different constraints of the tested functions and software components. So, in short, an automated testing tool slightly modifies the parameters of a function to check if it deviates from the expected behaviour \cite{Myers2012}. In another pertinent research field, adversarial machine learning, a similarity can be noticed: an adversarial attack method creates slight modifications in data features to deceive a model, causing misclassifications and unexpected behaviors \cite{Vitorino2023a}. Therefore, both automated testing tools and adversarial learning methods serve similar purposes and there could be significant benefits to their integration.

Due to the lack of publications addressing the use of adversarial learning methods to improve software testing tools, with data perturbations generated according to the specific constraints of the tested functions and software components, this work investigates the recent scientific developments and technological advances of these fields. The main research question of this literature review was: \textit{How can automated software testing tools be improved with constrained adversarial machine learning methods?}.

To guide the research performed in the scope of each field, software testing and adversarial learning, the main question was divided into two narrower sub-questions:

\begin{description}
\item[RQ1:] How can adversarial learning methods be leveraged for constrained data generation?
\item[RQ2:] How can automated software testing tools benefit from constrained data generation?
\end{description}

By investigating the current approaches in a methodical manner, this work can guide researchers and developers to enhance automated testing tools with adversarial learning methods and improve the resilience and robustness of the System Under Test (SUT). This paper is organized into multiple sections. Section \ref{sec:metho} describes the search methodology. Section \ref{sec:findrq1} and Section \ref{sec:findrq2} present and discuss the found publications for RQ1 and RQ2. Finally, Section \ref{sec:con} presents the concluding remarks and future research topics.

%% This is file 'section-methodology.tex'
\section{Search Methodology}
\label{sec:metho}

This section describes the methodical review process employed for both RQ1 and RQ2. To achieve a transparent, replicable, and complete literature review, a part of the PRISMA methodology \cite{Moher2015} was followed. Search terms were created to be used in reputable bibliographic databases, and several inclusion and exclusion criteria were defined. After an initial screening phase with the titles and abstracts of the found publications, their full texts were assessed for eligibility, and only then they were included in the review.

\subsection{Search Terms}

Search terms were chosen after a careful initial analysis of the literature. Since constrained data generation is a relatively unexplored research topic without standardized keywords, the search had to cover broader terms to prevent narrowing it down too much and obtain relevant publications for the scope of the review. Therefore, some word variations were also considered to widen the search to more common concepts like \verb|number generators| and \verb|conditional generators|, as well as approaches that address \verb|constraints| and \verb|restrictions|, which are word variations represented by "*".

Considering that RQ1 and RQ2 share a common scope, constrained data generation, only their target field was different in the search queries. For RQ1, the query included the \verb|adversarial| term to focus on approaches related to adversarial machine learning and adversarial perturbations. For RQ2, the target field was \verb|software testing| to analyze the publications specific to this field. Table \ref{tab:queryrq1} presents the combined query, where the terms of each scope were aggregated with AND operators.

\renewcommand\theadfont{}
\renewcommand{\arraystretch}{1.2}

\begin{longtable}{@{} m{0.19\textwidth} m{0.76\textwidth} @{}}
\caption{Search query for RQ1.}
\label{tab:queryrq1} \\
\toprule
Scope & Keywords \\
\midrule
Constrained & \textbf{constrained}* OR \textbf{conditional}* OR \textbf{restricted}* \\
Field & \textbf{adversarial learning} OR \textbf{software testing} \\
Data & \textbf{data} OR \textbf{sample} OR \textbf{example} OR \textbf{number} \\
Generation & \textbf{generation} OR \textbf{generator} \\
\bottomrule
\end{longtable}

\subsection{Search Sources}

The primary search source was Science Direct \cite{SourceElsevier}, which is a large bibliographic database of scientific journals and conference proceedings provided by the publisher Elsevier. Due to their acknowledged relevance for scientific literature of software engineering, the search also included the digital libraries of the Institute of Electrical and Electronics Engineers (IEEE) \cite{SourceIEEE}, the Association for Computing Machinery (ACM) \cite{SourceACM},  and the Multidisciplinary Digital Publishing Institute (MDPI) \cite{SourceMDPI}.

\subsection{Inclusion and Exclusion Criteria}

The review was performed in the beginning of 2024 and was focused on peer-reviewed conference papers and journal articles introducing or applying constrained generation approaches, in the English language. Considering that both adversarial learning and software testing are active areas of research and had relevant technological advances in recent years, the selected time frame was 2017 to 2023, the latest 7 whole years of scientific developments, at the time the review was started. To limit the findings to recent developments published and stored in these databases, backward snowballing was not performed in this review.

The widened search queries led to more relevant publications within this time frame, but also to many publications that were not aligned with the purpose of this review. To screen the publications and assess their eligibility, several criteria were defined. These intended to include only the publications presenting possible approaches that took constraints into account, excluding general surveys and reviews that were not focused on constraints.

Furthermore, the criteria also excluded publications that mentioned constraints or conditions but not during their data generation processes, and publications that did not detail how the utilized constraints were applied nor how they affected the data generation. Table \ref{tab:incexc} provides an overview of the defined inclusion and exclusion criteria for both RQ1 and RQ2.

\renewcommand\theadfont{}
\renewcommand{\arraystretch}{1.2}

\begin{longtable}{@{} m{0.5\textwidth} m{0.46\textwidth} @{}}
\caption{Inclusion and exclusion criteria.}
\label{tab:incexc} \\
\toprule
Inclusion Criteria & Exclusion Criteria \\
\midrule
\shortstack[l]{ \\
    IC1 Peer-reviewed journal article \\ or conference paper \\
    IC2 Published from 2017 to 2023 \\
    IC3 Available in the English language \\
    IC4 Introduced or applied an approach \\ for constrained data generation}
& \shortstack[l]{
    EC1 Duplicated publication \\
    EC2 Survey or review \\
    EC3 Constraints not in generation \\
    EC4 Constraints not detailed \\
    EC5 Full text not available} \\
\bottomrule
\end{longtable}

%% This is file 'section-findings-adversarial.tex'
\section{Adversarial Machine Learning}
\label{sec:findrq1}

\begin{figure}[ht!]
    \centering
    \vspace{5mm}
    \includegraphics[scale=0.7]{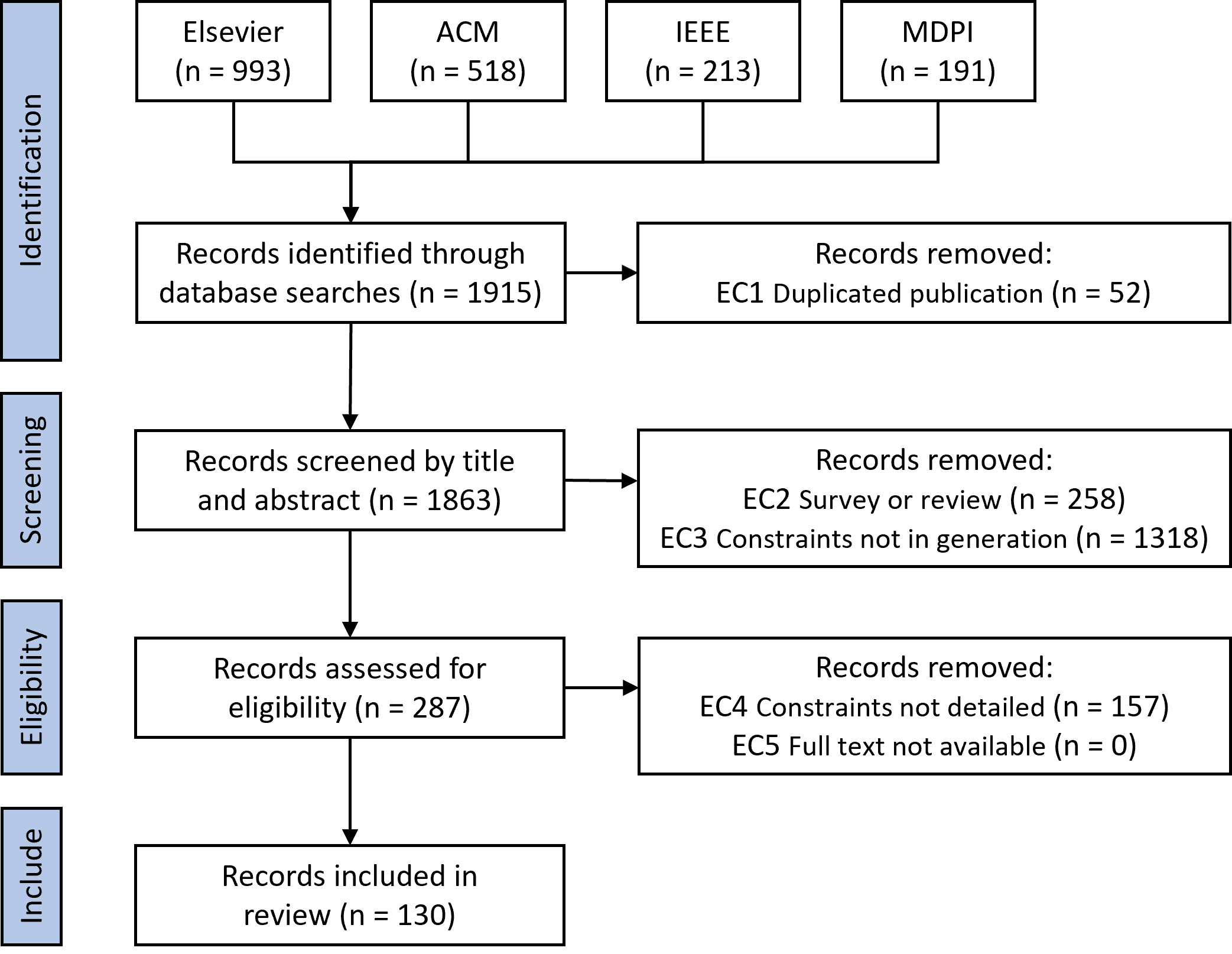}
    \caption{PRISMA search process for RQ1.}
    \label{fig:prismarq1}
\end{figure}

This section presents and discusses the findings of RQ1, systematizing the constrained data generation applications for adversarial machine learning. A total of 1915 records were initially obtained by applying the query to the contents of the publications stored in the selected databases. After the screening and eligibility assessment phases, 130 publications were included in the review. This process is detailed in Figure \ref{fig:prismarq1}.

The included publications were systematized to identify the current applications of constrained data generation and the methods they are based on. A total of 98 applications were identified across 11 sectors and industries, although a few were more general and not specific to a single industry. The number of applications grew from 2 in 2017 to 36 in 2023, with the cybersecurity and healthcare sectors being the most prominent and having more diverse applications, followed by the aerospace and energy sectors. These are the key sectors where it was identified that constrained data generation is necessary and there is a growing number of applications.

Table \ref{tab:adv} summarizes the reviewed applications and the methods they were based on. These base methods were either explicitly built upon or their concepts were implicitly used in new implementations. The table is divided according to the main sectors and industries: aerospace, automotive, chemical, cybersecurity, economy, energy, geological, healthcare, mechanical, music, and water. The remaining publications that were more broad are included in a general subdivision at the end of the table.

\newpage

\renewcommand\theadfont{}
\renewcommand{\arraystretch}{1.2}
\setlength{\tabcolsep}{10pt}

\begin{longtable}{@{} m{0.18\textwidth} m{0.56\textwidth} m{0.16\textwidth} @{}}
\caption{Applications of constrained data generation.}
\label{tab:adv} \\
\toprule
\shortstack{Sector / \\ Industry} & Applications & \shortstack{Base \\ Methods} \\
\midrule
\endfirsthead
\toprule
\shortstack{Sector / \\ Industry} & Applications & \shortstack{Base \\ Methods} \\
\midrule
\endhead
\multirow{9}{*}{Aerospace}
    & Synthetic-Aperture Radar (SAR) image reconstruction, and hyperspectral classification
    \cite{FuentesReyes2019,Wang2020b,Chen2021,Costa2021} & CGAN, CVAE \\ \cline{2-3}
    & Satellite mapping of surface water, road surface area, reflectance, and radiance
   \cite{Kim2019,Han2020,Mizuochi2021,Cira2022} & CGAN \\ \cline{2-3}
    & Precipitation estimation, aircraft detection, remote sensing, and signal separation
    \cite{Zhang2018,Hayatbini2019,Pan2021,Heller-Kaikov2023} & CGAN \\ \cline{2-3}
    & Object detection in unmanned drone inspection tasks with Unmanned Aerial Vehicles (UAVs)
    \cite{Chen2021b,Munawar2022} & CGAN \\
\midrule
\multirow{7}{*}{Automotive}
    & Urban traffic trajectory simulation, and parking occupancy estimation
    \cite{Zhang2020a,Cheng2021} & CGAN, CVAE \\ \cline{2-3}
    & Motor vehicle transmission gear and Inertial Measurement Unit (IMU) signal analysis
    \cite{Jaafer2020,Li2021b} & CGAN \\ \cline{2-3}
    & Ultrasonic signal sensors for autonomous driving systems
    \cite{Popperli2019} & CGAN \\ \cline{2-3}
    & Vehicular and transportation network simulation
    \cite{Falahatraftar2021,Kocayusufoglu2022} & CGAN \\
\midrule
\multirow{4}{*}{Chemical}
    & Chemical production process fault detection and diagnosis
    \cite{Zhu2021,Qin2022a} & CGAN, WGAN \\ \cline{2-3}
    & Soft sensors for multiple chemical compound analysis
    \cite{He2022} & CGAN, WGAN \\ \cline{2-3}
    & Environmental feature extraction in aerosol optical depth sensors
    \cite{Zhang2021} & CGAN \\
\midrule
\multirow{11}{*}{Cybersecurity}
    & Network intrusion detection, anomaly detection, and cyber-attack classification
    \cite{Alhajjar2021,Chernikova2022,Dina2022,Vitorino2022,Wang2022,Kumar2023a} & CGAN, evolutionary algorithms \\ \cline{2-3}
    & Illegal webpage detection, and malicious domain name detection
    \cite{Chernikova2022,Wan2020,Liu2021a} & CGAN, evolutionary algorithms \\ \cline{2-3}
    & Input generation for software testing, and malware detection in code blocks
    \cite{Gao2020,Guo2022,Kasarapu2022,Amrith2023} & CGAN, fuzzy algorithms \\ \cline{2-3}
    & Fake user detection, and behavior and keystroke dynamics generation
    \cite{Chonwiharnphan2020,Esmaili2021,Eizagirre2023} & CGAN \\ \cline{2-3}
    & Speech recognition and denoising, and voice spoofing detection
    \cite{Qian2019,Chen2020,Ram2020,Ding2021} & CGAN \\ \cline{2-3}
    & Data masking, obfuscation, anonymization, and compression
    \cite{Yoon2020,Liu2021,Khwaja2022,Sun2023a} & CGAN \\
\midrule
\multirow{3}{*}{Economy}
    & Monte Carlo simulation design for economic studies
    \cite{Athey2021} & CGAN, WGAN \\ \cline{2-3}
    & Multi-period financial portfolio time-series simulation
    \cite{Sun2023b} & CGAN \\
\midrule
\multirow{4}{*}{Energy}
    & Power distribution, economic load dispatch, and optimal power flow simulation
    \cite{Kocayusufoglu2022,Fang2021,Li2021,Song2023,Zeng2023,Yin2023} & CGAN, graph algorithms \\ \cline{2-3}
    & Anomalous energy consumption detection, and electricity theft detection
    \cite{Gong2020,Liu2020} & CVAE, WGAN \\ \cline{2-3}
    & Charging behavior, and supply and demand forecasting
    \cite{Moon2020,Qin2022,Zhang2022d,Zhang2022a,Bu2023,Ye2023} & CGAN \\
\midrule
\multirow{3}{*}{Geological}
    & Seismic data, sedimentary facies, and geological patterns simulation
    \cite{Chang2021,Hu2023,Xu2023,Sun2023} & CGAN \\ \cline{2-3}
    & Virtual landscape and terrain authoring
    \cite{Guerin2017,Zhang2022b} & CGAN \\
\midrule
\multirow{18}{*}{Healthcare}
    & Magnetic Resonance Imaging (MRI) and Positron Emission Tomography (PET) analysis
    \cite{Silva2019,Teixeira2021,Amirrajab2022,Qiang2022,Jafaritadi2023,Jung2023} & CGAN, WGAN \\ \cline{2-3}
    & X-ray and Computed Tomography (CT) scan analysis
    \cite{Liu2022b,Jin2023,GarciaHernandez2023,Mendes2023,Li2023} & CGAN, WGAN \\ \cline{2-3}
    & Electrocardiogram (ECG) and Electroencephalogram (EEG) signal analysis
    \cite{Hagad2021,Zhou2021,Karabulut2022,Xia2023,Qu2023,Biswas2023} & CGAN, WGAN \\ \cline{2-3}
    & Fundus image generation and Retinopathy screening
    \cite{Xie2023} & CGAN \\ \cline{2-3}
    & Electrodermal Activity (EDA) generation for stress detection
    \cite{Ehrhart2022} & CGAN \\ \cline{2-3}
    & Lung and skin lesions detection
    \cite{Havaei2021} & CGAN \\ \cline{2-3}
    & Mammographic density segmentation and breast cancer diagnosis
    \cite{Saffari2020,Gur2022,Zhang2022,Inan2023} & CGAN \\ \cline{2-3}
    & Gene expression profiling for cancer diagnosis
    \cite{Wu2023} & CGAN \\ \cline{2-3}
    & Microscopic blood smear tests and single-cell segmentation
    \cite{Bailo2019,Tasnadi2023} & CGAN \\ \cline{2-3}
    & Vocal cord and voice disorder detection
    \cite{Chui2020} & CGAN \\ \cline{2-3}
    & Connectomes segmentation for reconstruction of neural circuits
    \cite{Chen2018} & CGAN \\
\midrule
\multirow{5}{*}{Mechanical}
    & Rotating machinery and rolling bearing fault diagnosis
    \cite{Yu2019,Dixit2021,Ahang2022,Li2022,Liu2022,Peng2022} & CGAN, CVAE, WGAN \\ \cline{2-3}
    & Industrial equipment failure, conveyor belt, and joint damage detection
    \cite{Wang2020a,Chen2022,Guo2022a,Chen2023,Zhao2023} & CGAN, CVAE, WGAN \\ \cline{2-3}
    & Wind turbine gearbox fault diagnosis
    \cite{Wang2020,Liu2022a,Zhang2022c} & CGAN, CVAE \\
\midrule
\multirow{3}{*}{Music}
    & Variable-length music and audio generation
    \cite{Haque2020,Li2021a} & CGAN, CVAE \\ \cline{2-3}
    & Cross-modal musical performance generation
    \cite{Chen2017} & CGAN \\
\midrule
\multirow{3}{*}{Water}
    & Water flow simulation for water supply and distribution systems
    \cite{Kocayusufoglu2022,Li2021,Rajabi2023} & CGAN, graph algorithms \\ \cline{2-3}
    & Maritime vessel trajectory prediction in waterways
    \cite{Jia2023} & CGAN \\
\midrule
\multirow{13}{*}{General}
    & Architectural space layout generation
    \cite{Aalaei2023} & CGAN, graph algorithms \\ \cline{2-3}
    & Motion, stretches, and compressions simulation
    \cite{Barsoum2018,Geng2020} & CGAN, WGAN, skinning algorithms \\ \cline{2-3}
    & Portrait picture generation according to age and facial attributes
    \cite{Li2018,Marzouk2019,Chen2021a} & CGAN \\ \cline{2-3}
    & Font generation according to different font styles
    \cite{Hassan2023} & CGAN \\ \cline{2-3}
    & Pathfinding for Very Large-Scale Integration (VLSI) of integrated circuits
    \cite{Utyamishev2023} & CGAN \\ \cline{2-3}
    & Wireless indoor signal denoising, positioning, and tracking
    \cite{Belmonte-Hernandez2020,Tang2023} & CGAN \\ \cline{2-3}
    & Word-gesture text input, handwritten text, and digit recognition
    \cite{Jahic2019,Kang2022,Chu2023} & CGAN \\
\bottomrule
\end{longtable}

The increasing use of constraints across various sectors and industries highlights that better data can be generated when the specificities of the task at hand are considered. Nonetheless, since most publications handle image classification datasets, they employ methods with simple constraints to generate new synthetic images according to the classes of a dataset. Most approaches are based on improved versions of the Generative Adversarial Network (GAN) \cite{Goodfellow2014} and the Variational Autoencoder (VAE) \cite{Kingma2014}, such as Conditional GAN (CGAN) \cite{Mirza2014}, Wasserstein GAN (WGAN) \cite{Arjovsky2017}, and Conditional VAE (CVAE) \cite{Sohn2015}.

Even though the CGAN and CVAE improved versions of deep learning algorithms spread across the literature because they can learn to generate different data for different classes, the samples of each class are freely generated without addressing any complex constraint. For more complex tasks in sectors that require other data types like tabular data and time series, some approaches are starting to rely on methods that support more rigorous configurations and more complex constraints. For instance, some researchers employ evolutionary computation with swarm intelligence and custom-built Genetic Algorithms (GAs) \cite{Alhajjar2021}, and others apply fuzzy logic in their algorithms to deal with the uncertainty of the information \cite{Gao2020}.

Despite the wide range of applications of constrained data generation, the few approaches that could potentially be transferable to automated software testing tools were mainly developed for cybersecurity applications related to communication networks and cyber-physical applications in energy grids and water distribution networks. Due to the low-quality data that conventional methods would generate in these complex domains, some researchers encapsulated the data generation processes in mechanisms that enforce task-specific constraints.

In \cite{Li2021}, the authors intended to simulate electrical grids and water supply and distribution systems. Each action or event in one part of these cyber-physical systems would affect the entire energy and water networks, so the presence of a given value at a given feature would restrict the values that other features could have. Linear equality constraints were defined to fulfill the physical capacity requirements of each network and multiple simulations were performed to validate them. This optimization approach could be useful to model the relationships between different nodes of a network and different software components, but the constraints would need to be carefully redesigned for each minor change in the tested functions or to be transferred to different SUT architectures. 

To provide a structure that could be adapted to changes, in \cite{Kocayusufoglu2022}, energy and water networks were addressed through graph theory. The graph structure provided a more rigorous representation of the electric and water flow physics between different nodes in a network, enabling the simulation of cyclic trends and acyclic congestions, as well as their effects in other nodes. Nodes and their connections could be added or removed to adapt the structure to different flow networks, although it was tailored to the specific physical constraints of those domains. Additionally, it is also relevant to highlight the mechanism introduced in \cite{Geng2020} to perform physically accurate simulations of stretches and compressions of objects in three-dimensional spaces. Despite not being as rigorous as flow graphs and only being tested with triangulated cloth meshes of human body poses and joint rotations, this mechanism could be transferable to the perturbation of images and objects provided to the SUT and test its complex code structure.

Regarding biomedical images, in \cite{Havaei2021}, the authors addressed the insufficient label granularity of most datasets. Since multiple sub-types of a disease can be aggregated into a single broader label, the data samples of a class can exhibit entirely different data distributions and feature correlations. Even if a CGAN is used to distinguish between different classes of a dataset, the generated data may mix the characteristics of multiple sub-types and therefore may not correspond to an actual disease. Conditioning vectors were used to restrict the modification of relevant disease features according to the values exhibited by other similar samples that were presumed to be of the same sub-type. Even though this mechanism is still based on a CGAN and would need to be redesigned for other data types, the concept of configuring class-specific value constraints could be useful to improve data quality and generate relevant parameter combinations in a reduced amount of time.

These sub-classes were further explored in \cite{Vitorino2022}, where the authors established that an adversarial example must be valid within its domain structure, which corresponds to the input of the software component being tested, and be coherent with the characteristics and purposes of each class. Constrained data perturbations were generated by analysing the value intervals of individual features and the value combinations of groups of features. Despite being developed for network intrusion detection, the method created constrained adversarial examples that fulfilled complex constraints. This method could be useful to software testing because the tested functions could be aggregated into classes according to their functionalities and function-specific value perturbations could be performed, quickly generating more adequate parameter combinations to achieve a high code coverage.

Overall, the reviewed adversarial machine learning approaches present valuable insights for automated software testing applications, although they are tailored to the specificities of cyber-physical systems. These mechanisms could be redesigned to consider the value constraints of the parameters of each tested function and generate high-quality data to test different software components. Therefore, it is also pertinent to explore the recent advances in automated software testing tools and analyze the ways that constrained data generation could be used to improve them.

%% This is file 'section-findings-software.tex'
\section{Automated Software Testing}
\label{sec:findrq2}

\begin{figure}[ht!]
    \centering
    \vspace{5mm}
    \includegraphics[scale=0.7]{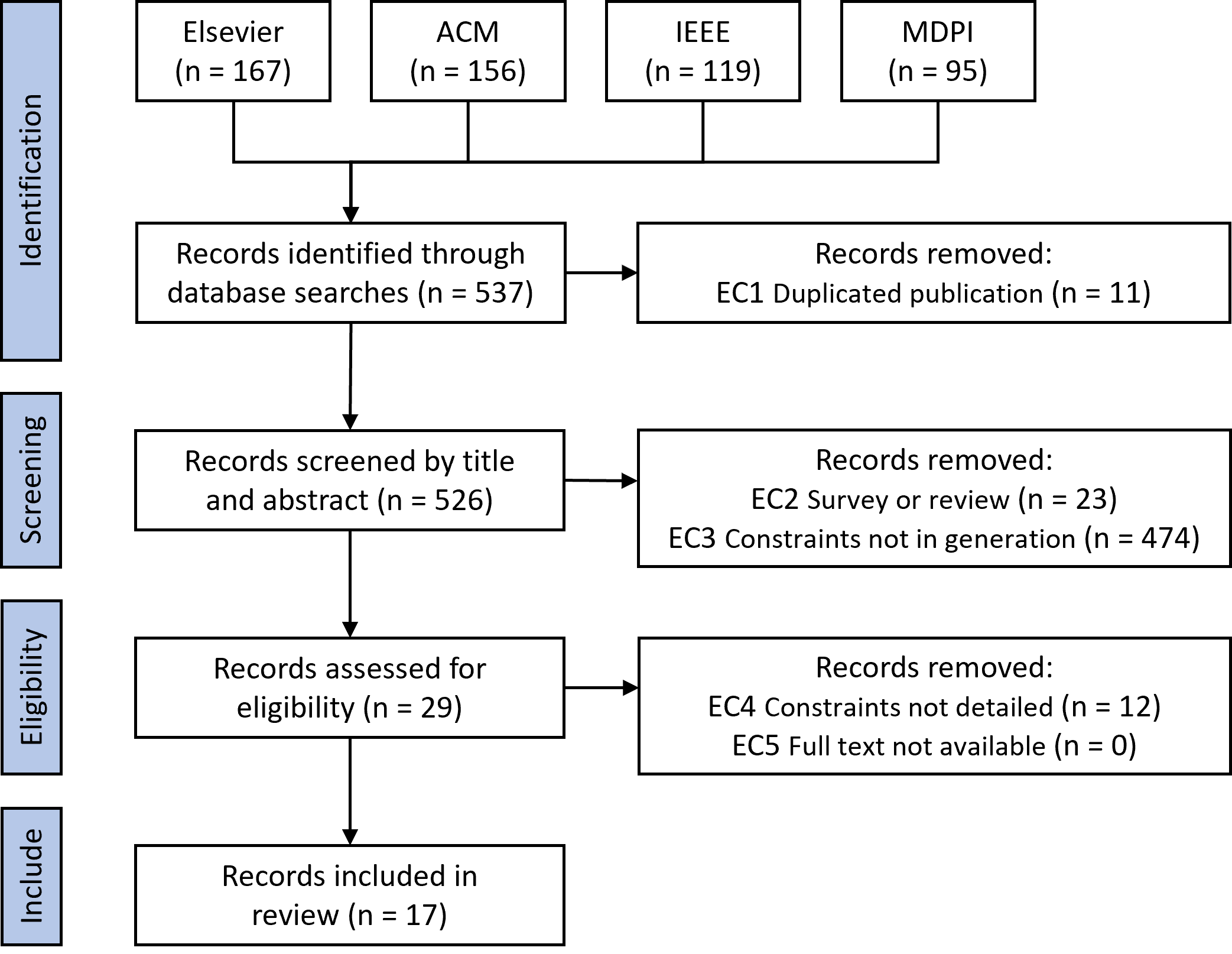}
    \caption{PRISMA search process for RQ2.}
    \label{fig:prismarq2}
\end{figure}

This section presents and discusses the findings of RQ2, thoroughly analyzing the advantages and limitations of approaches specific to software testing. A total of 537 records were identified in the selected databases. After the screening of their titles and abstracts and the assessment of their full texts for eligibility, 17 publications were included in the review. This process is detailed in Figure \ref{fig:prismarq2}.

Software testing approaches can be divided into a three-fold: (i) white-box, (ii) black-box, and (iii) grey-box. The first is mostly executed from a developer's point of view, where all the domain constraints and the system’s inner workings are known and leveraged. The black-box approach is characterized by its lack of knowledge and access to the source code~\cite{Martin-Lopez2021a}. In a black-box testing setting, the tester has only information about what the system is supposed to do and how it can be interacted with. However, further intricacies of the SUT are unknown~\cite{Martin-Lopez2021a}. Lastly, grey-box testing combines both approaches. However, the implementation of the SUT and its domain is only partially known~\cite{Rajamanickam2019}.

Software testing automation can be extremely challenging, depending on the approach and levels of testing considered. The key aspect to achieve automation relies on the construction of the test cases, which are a reflection of the SUT's functions parameter values and their execution sequence \cite{Zheng2021}. When considering the generation of input values to test the system, one must deal with the difficulties related with finding the right input values along with their combination and the right execution sequence, since the prior execution of system functionality may alter its state, which leads to different system execution paths \cite{Zheng2021}.

Depending on the information available to the automated testing tool these can have enormous search spaces, which is not ideal when considering the possible execution time \cite{Khan2012}. Depending on the software and the levels of testing that the automated tool is trying to achieve, a white-box automated testing tool with access to more domain information might perform better and more efficient testing of the SUT's domain. Nonetheless, testing a system from a user perspective using black-box techniques may sometimes raise different exceptions and errors that are not usually expected nor visually understandable in the source code \cite{Anwar2019}.

Considering the different domain constraints that different systems have, this review has shown that although random generation of data may be capable of achieving reasonable test path coverage of simpler paths, it is highly unpractical on more complex ones. Complex functionalities usually contain more dependencies, making the combination of the right parameters extremely complex. Moreover, the literature shows that in a black-box context, the input space for certain data types is continuous, and therefore different techniques such as providing boundary values\cite{Martin-Lopez2021,Martin-Lopez2020,Haraldsson2017}, leveraging system feedback \cite{Martin-Lopez2021, Walkinshaw2017, Kim2018} and applying search techniques can reduce computational complexity \cite{Biagiola2019}.

In a grey-box setting, the authors have also attempted to reduce the input space by resorting to specifications that reflect source code knowledge to constraint the input \cite{Alazzawi2019} and techniques for quick exploration of the input space \cite{Huang2017}. In a white-box setting, automated testing can be interpreted as a search problem, with most existing methods focusing on metaheuristics and their optimization to find more diverse test cases \cite{Esnaashari2021,Avdeenko2021, Zhu2019, Yao2020, Jaiswal2021, Dang2020}. One publication has attempted to use GANs for software testing \cite{Guo2022}, but like many white-box approaches, it only focuses on numerical input generation, which may be insufficient to fully test all the inputs of modern software applications and information systems.

The gathered publications were analyzed to identify the test scope, the main input generation methods, the covered data types and the utilized data sources. This information is summarized in Table \ref{tab:soft}. The keyword "All" refers to numerical, textual, byte, boolean, and composite data types.

Regarding the methods for input generation from a black-box testing perspective, since there is no knowledge regarding the SUT, authors have found ways of constraining the input generation using other available data. For instance, Martin-Lopez et al. \cite{Martin-Lopez2021} decided to test a Representational State Transfer (REST) Application Programming Interface (API). Constrained input was produced via three different methods: (i) OpenAPI Specifications (OAS), which establishes the baseline information to interact with an API, and may include request examples, (ii) custom data generators which produce constrained input, and (iii) public knowledge bases. Similarly, in \cite{Haraldsson2017}, the authors propose using the data collected from a public API and mutate faulty input to produce test cases. Even though both approaches are capable of producing realistic test data, they rely on publicly available data, which might not be sufficient to cover the entirety of the SUT test paths.

Recent works have also considered the use of model-based approaches to solve the search problem by producing a model of the SUT capable of generating constrained input. For instance, Martin-Lopez \cite{Martin-Lopez2020} resorts to three different methods for testing web APIs: (i) search-based methods, (ii) mutating API JSON input and output, and (iii) computing metamorphic testing. The goal of the authors is to not diverge from the initial input data, preserving its quality by using artificial intelligence and generative algorithms to produce constrained test input. Walkinshaw et al. \cite{Walkinshaw2017} propose a method that combines Learning-based Testing (LBT) and Query By Committee (QBC). QBC is used in this context to circumvent LBT’s dependence between model inference algorithm and the test-generation algorithm. This mechanism allows LBT to select inputs based on the combined uncertainty of the models, inferring a behavioural model of the SUT and selecting the test that it is least certain about.

\newpage

\renewcommand\theadfont{}
\renewcommand{\arraystretch}{1.2}
\setlength{\tabcolsep}{10pt}

\begin{longtable}{@{} m{0.12\textwidth} m{0.34\textwidth} m{0.10\textwidth} m{0.29\textwidth} @{}}
\caption{Characteristics of constrained testing methods.}
\label{tab:soft} \\
\toprule
\shortstack{Test \\ Scope} & Method & \shortstack{Data \\ Types} & Data Sources \\
\midrule
\endfirsthead
\toprule
\shortstack{Test \\ Scope} & Method & \shortstack{Data \\ Types} & Data Sources \\
\midrule
\endhead
\multirow{7}{*}{Black-box} 
    & Custom Data Generators \cite{Martin-Lopez2021} & All & OpenAPI specifications and knowledge \\ \cline{2-4}
    & Metamorphic and Search-based Testing \cite{Martin-Lopez2020} & All & Input/Output data and API configuration \\ \cline{2-4}
    & Custom Mutators \cite{Haraldsson2017} & All & Faulty real user input \\ \cline{2-4}
    & Test By Committee \cite{Walkinshaw2017} & All & Custom specifications \\ \cline{2-4}
    & Reinforcement Learning \cite{Kim2018} & Numerical & Random data \\ \cline{2-4}
    & Input Distance \cite{Biagiola2019} & Numerical & Navigational model \\
\midrule
\multirow{4}{*}{Grey-box} 
    & Pairwise hybrid Artificial Bee Colony \cite{Alazzawi2019} & Numerical & Constraint specifications \\ \cline{2-4}
    & Locally Spreading, Points Algorithm, and Adaptive Random Test \cite{Huang2017} & Numerical & Defect-free confidence interval \\
\midrule
\multirow{10}{*}{White-box} 
    & Genetic Algorithm and Reinforcement Learning \cite{Esnaashari2021} & Numerical & Control flow graph \\ \cline{2-4}
    & Genetic Algorithm \cite{Avdeenko2021, Zhu2019, Yao2020} & Numerical & Random data \\ \cline{2-4}
    & Particle Swarm Optimization \cite{Jaiswal2021} & Numerical & Control flow graph \\ \cline{2-4}
    & D-Algorithm \cite{Zhang2019} & Numerical & Control flow graph and define-use chains \\ \cline{2-4}
    & Multi-objective Optimization Algorithm \cite{Feng2021} & Numerical & Random data \\ \cline{2-4}
    & Coevolutionary Genetic Algorithm \cite{Dang2020} & Numerical & Random data \\ \cline{2-4}
    & Generative Adversarial Networks \cite{Guo2022} & Numerical & Gcov file\\
\bottomrule
\end{longtable}

Additionally, Kim et al. \cite{Kim2018} reinterpreted this search problem as a reinforcement learning one. Despite achieving 100\% branch coverage during testing, for unforeseen arbitrary functions it only achieved 60.06\% branch coverage. Moreover, the algorithm only considers numeric input which depending on the SUT may render this approach unpractical for software testing. These approaches are as good as the amount of knowledge one has of the SUT, but as modern software and digital systems grow in size and complexity, so do the domain constraints and the number of branches that should be included in the configurations.

M. Biagiola et al. \cite{Biagiola2019} produce a navigational model of a web application by crawling it to discover possible test cases. The presented method’s first test is generated randomly and is added to a set of executed tests. The subsequent tests and concrete input vectors are selected depending on the distance between each candidate and the current set of executed test cases, where only the farthest case is computed. Their goal is to generate a set of test cases that diversify the coverage of the navigational graph. Even though they present a method that does not rely on exhaustive testing of the API that requires in-browser executions, they still rely on random input which dictates how long the system will take to find the next valuable input data to cover different test paths, since it is calculated by the distance formula. From a grey-box perspective, Alazzawi et al. \cite{Alazzawi2019} also tackle the same problem using a Pairwise hybrid Artificial Bee Colony algorithm, which takes random input and a configuration file with the constraints to find the most diversified test cases. However, depending on the manual configuration, the test cases can lack quality and diversity.

In a grey-box setting, Huang et al. \cite{Huang2017} tackle the broad input space and constraint it by resorting to the combination of a Locally Spreading Points algorithm along with a code defect-free confidence interval. The algorithm receives multiple generated test cases and evolves them to improve the minimum distance between points, achieving a better spread set of test cases. The approach improves effectiveness when used as an add-on to Adaptive Random Testing (ART). However, the method requires manual labeling of the test cases, regarding the confidence interval.

Even though black-box and grey-box methods can be quite useful to test a system regardless of its characteristics, the findings show that most research works attempt to automate software testing from a white-box perspective interpreting input generation as a search-based problem and resorting to metaheuristics to perform the search. As such many authors have also used GA to constraint the test cases generation~\cite{Esnaashari2021, Avdeenko2021, Zhu2019, Yao2020}. Esnaashari et al. \cite{Esnaashari2021} use the GA to solve the search-based along with reinforcement learning as a local search step within the GA. Their work is capable of achieving results faster, but it does not produce better coverage than other methods. In particular, Avdenkoo et al.~\cite{Avdeenko2021} and Zhu et al.~\cite{Zhu2019} focus on improving the genetic operators. Zhu et al. improve: (i) selection by resorting to symbolic execution technique to select test cases with more useful heuristic information, (ii) crossover by analyzing the coverage of the combination of test cases, and (iii) mutation by enforcing the modification of values according to specific constraints.

On the other hand, Avdeenko et al. focused on improving the fitness function of the algorithm and were able to achieve 100\% code coverage after multiple generations. Despite the quality of their work, Zhu et al. only consider the generation of numeric input in an extremely simple code sample. Avdeenko et al. do not provide much detail regarding the generated input, but the tool was tested in a more complex code structure. Yao et al. \cite{Yao2020} tackled the problem that is testing software with randomness and used GA along with a mathematical model capable of generating data according to a certain criterion, by observing the impact of random behavior on the software when given random input numbers and modifying them. In the white-box testing context, GAs are very popular. However, most approaches rely on generating random input and evolve it by testing iteratively the new input, using metrics such as the code coverage as fitness. Depending on the size of the software that is being tested, this approach may be just as unfeasible as the random approach.

Other authors have also approached the problem without necessarily resorting to the most traditional metaheuristic methods. For instance, Jaiswal et al. \cite{Jaiswal2021} decided to approach the problem using particle swarm optimization algorithm along with an improved fitness function that uses Control Flow Graph (CFG) to generate constrained input. This technique led to authors to achieve 100\% code coverage in only two iterations. However, the method only considers numeric input and the code analysed lacks complexity. Therefore, further tests should be made to understand if this is indeed a viable solution for constrained input generation. Zhang et al. \cite{Zhang2019} propose a version of the D-algorithm for software capable of constraining the input generation process to directed local search. Their method requires a CFG of the code, as well as define-use chains, which represent where a variable is defined in the node and when it is used in the CFG.

Feng et al. \cite{Feng2021} developed their own multi-objective optimization algorithm for path coverage-oriented test data generation. They start by randomly generating data and use it to discover the easy-to-cover paths. Then they perform mutations on the input data generated to discover the more complex paths. Dang et al. \cite{Dang2020} attempt to automate and reduce the search domain in the context of mutation testing, which is a fault-oriented software testing technique. Their work consists of a method for generating test data applying a dimensionality reduction on the search domain based on a mutant stubbornness. The authors approach this problem as an optimization one and decide to use a coevolutionary GA to perform the task of generating test data to kill mutants.

Guo et al. \cite{Guo2022} took one step further and decided to use WGAN + Gradient Penalty (WGAN-GP) to produce constrained test data input. The authors focus on performing unit and integration testing. The WGAN-GP is trained on the execution path information to learn the behaviour of the SUT. The trained GAN is then capable of producing and selecting new test data to increase branch coverage. The GAN is trained with the execution path of test input using the Gcov tool. Gcov is a source code coverage analysis tool capable of counting the number of times each statement is a program is executed. The algorithm then produces new data that is used along with Gcov to generate execution path information, which is then fed into the GAN again. This way, the authors are capable of maximizing branch coverage. Compared to random testing, their work was capable of improving test coverage of 5 out of 7 functions in the unit testing.

Overall, the white-box testing approaches mostly resort to heuristics and perform iterative testing of the generated input against the SUT to check if the coverage increases. Exhaustive evaluation of all possibilities should be avoided, for it is very expensive when considering the complexity of modern systems. From a grey-box and black-box perspective, the approaches usually lack knowledge regarding the SUT and its constraints, which can lead to more time-consuming and resource-intensive testing. Automated testing tools could leverage adversarial methods like GANs capable of constraining the input more efficiently and reducing the search space. By testing specific parameter combinations tailored to the characteristics of each function, constrained adversarial data generation approaches could present significant improvements to these tools. Therefore, it is pertinent to perform further research and develop new methods and tools to improve the quality of the data utilized in automated software testing tools.

%% This is file 'section-conclusions.tex'
\section{Conclusions}
\label{sec:con}

This work reviewed the recent scientific advances of constrained data generation in a methodical manner. The state-of-the-art methods for constrained data generation in the adversarial machine learning and automated software testing fields were analyzed, identifying research gaps and opportunities to transfer knowledge from adversarial attacks to enhance testing tools and improve the resilience and robustness of information systems.

Regarding the current testing tools, there are several challenges that hinder full automation. Recent works attempt to constraint the input that is generated by analyzing the SUT empirically or resorting to specifications, although most perform exhaustive search by mutating parameters in order to find different branches of the SUT that are yet to be tested, which is unpractical. This suggests it may be difficult to use adversarial learning for black-box and grey-box testing, which have more complex constraints that are not directly provided to the tool.

Since white-box testing is commonly performed during the development of a system to make it more secure, when developers have knowledge of the utilized APIs and architectures, this is the key aspect where constrained adversarial learning could be valuable. Even though most approaches were developed for general applications, they can potentially be adapted to generate high-quality data according to the constraints of the SUT. Adversarial examples could be used to further improve testing tools with different parameter combinations specifically crafted to deceive the tested functions and software components, reducing the time and the computational resources required to achieve a high test coverage in complex systems.

In the future, the use of adversarial machine learning methods and other artificial intelligence techniques to enhance software development and testing should be further explored. For instance, to perform white-box unit testing, natural language processing could be leveraged to analyze the code and extract its constraints to quickly adapt the adversarial attack methods for the relevant test cases and improve computer code coverage in a more efficient way. From a black-box perspective, it may be harder to have a fully accurate code coverage analysis, but fuzzy logic could be combined with adversarial reinforcement learning as an API behaviour learning technique to generate faulty test input according to the characteristics of different APIs, providing more adaptive software testing tools.

\newpage

% \paragraph{\textbf{Ethical approval and Consent to participate}}
% \hfill \break

% Not applicable.

% \paragraph{\textbf{Consent for publication}}
% \hfill \break

% Not applicable.

% \paragraph{\textbf{Availability of supporting data}}
% \hfill \break

% Not applicable.

% \paragraph{\textbf{Conflicts of Interest}}
% \hfill \break

% The authors declare no conflict of interest.

\paragraph{\textbf{Author Contributions}}
\hfill \break

Conceptualization, J.V. and I.P.; methodology, J.V., T.D. and T.F.; validation, E.M. and I.P.; investigation, J.V., T.D. and T.F.; writing, J.V., T.D. and T.F.; supervision, E.M.; project administration, I.P.; funding acquisition, I.P. All authors have read and agreed to the published version of the manuscript.

\paragraph{\textbf{Funding}}
\hfill \break

This work was done and funded in the scope of the BEHAVIOR project (NORTE2030-FEDER-00576300 no. 14391). This work was also supported by UIDB/00760/2020.

% This work was supported by the UIDB/00760/2020 and UIDP/00760/2020 projects.

% \paragraph{\textbf{List of Abbreviations}}
% \hfill \break

% API - Application Programming Interface

% CGAN - Conditional GAN

% CVAE - Conditional VAE

% GA - Genetic Algorithm

% GAN - Generative Adversarial Network

% REST - Representational State Transfer

% SUT - System Under Test

% VAE - Variational Autoencoder

% WGAN - Wasserstein GAN

% \newpage

%% If you have bibdatabase file and want bibtex
%% to generate the bibitems, please use:
\bibliographystyle{elsarticle-num} 
\bibliography{thebibliography}

%% else use the following coding to input the 
%% bibitems directly in the TeX file:
%% \begin{thebibliography}{00}
%% \end{thebibliography}

\end{document}